\documentstyle[12pt]{article}
\author{Hans - J\"urgen Schmidt}
\title{How to measure spatial distances?}
\date{}
\begin{document}
\maketitle

\bigskip

\centerline{
Universit\"at  Potsdam, Institut f\"ur Mathematik}
\centerline{  Projektgruppe
Kosmologie}
\centerline{
      D-14415 POTSDAM, PF 601553, Am Neuen Palais 10, Germany}

\begin{abstract}
The use of time--like geodesics to measure temporal
distances is better justified than the use of
space--like geodesics for a measurement of spatial
distances. We give examples where a ''spatial distance''
 cannot be appropriately determined by the
length of a space--like geodesic.

\end{abstract}

\section{Introduction}

Let us consider two space--time points (events) which can be
connected by a time--like line. What is the distance between
them? The most common answer is as follows: connect these
points by a time--like geodesic; the natural length of
 this \footnote{To discuss what to do if this
fails to be unique is beyond the scope of this letter.}
 geodetic segment gives the desired temporal distance.

This answer becomes plausible by imagining a test particle
 (freely falling rocket) measuring its elapsed eigentime.

	Now, let us consider two other events which can be
connected by a space--like line and ask the same question.
It is tempting to transform the above mentioned most common
answer also to this case, as done e.g. in ref. [1];
 but then already the correspondingly transformed
 plausibility argument would require the
introduction of tachyons.

It is the aim of the present letter to discuss
the headline--question both from geometric and
from physical points of view.

\section{The geometric point of view}

Let us restrict to the class of smooth
( $= \  C^{\infty}$) space--times $V_4$ which are
globally hyperbolic, oriented and time--oriented.
These assumptions already exclude the existence of
closed time--like curves \footnote{We consider
smooth curves only}, cf. [2], whereas closed
space--like curves always exist in such space--times.

Supposed, we would exclude space--times which
possess closed space--like geodesics, then e.g. all the
closed Friedmann models would be excluded, and this
situation  we do not want. So already at this level,
space--like and time--like curves are qualitatively
different objects, there is no duality between them.

This difference appears already in the Minkowski
space--time $M_4$ of Special
Relativity Theory, let us give two well--known
examples:

 1. Let two events of $M_4$
 be connected by a time--like
curve, then they can also be connected by a
space--like curve.

 2. Let two events of $M_4$
 be connected by a time--like
curve, then they can also be connected by a
 time--like geodesic.

For both these statements it holds: If we
simultaneously interchange
''space--'' and ''time--'', then both become wrong.

\section{Geodesics in the de Sitter space--time}

To prevent the extra--discussion connected with
singular points we consider now the de Sitter
space--time represented as closed Friedmann universe.
The metric reads

\begin{equation}
ds^2 \ = \ dt^2  \ - \ \cosh^2t[dr^2 \ + \
\sin^2r(d\psi^2 +\sin^2\psi \ d\phi^2)]
\end{equation}

\noindent
It is  a connected simply connected
smooth geodesically complete space--time.
It holds:

1. If two events can be connected by a time--like
curve, then  they can also be connected by a
time--like geodesic; and this geodesic is unique.

2. If two events can neither be connected by a time--like
nor by a light--like curve, then they can be connected
by a space--like curve; however, in general they cannot
be connected by a space--like geodesic.

\bigskip

This last fact was already known to de Sitter
himself, cf. [2] and the more detailed calculation
in [3]. Here we want to give a new and more
geometric proof of that statement. To this end we fix any
point $p$ in the de Sitter space--time (1); all space--like
geodesics starting at $p$ will intersect again after
eigendistance $\pi$ in the antipodal point $q$.
Let $U$ be a small open neighbourhood around $q$
and let $V$ be the interior of the future
light--cone of $q$. Fix any $r \ \in \ U \ \cap \ V$, then
$p$ and $r$ cannot be connected by a space--like geodesic,
because, due to construction, all space--like geodesics
starting at $p$ end up in $q$ coming from a space--like
direction.

\bigskip

The situation is as follows: Nobody will have doubts that
the spatial distance between $p$ and $q$ equals
$\pi$, one half of the circumference of a circle
with unit radius. Now, $r$ is arbitrarily close to $q$,
so the spatial distance between $p$ and $r$
should be defined and should be very near to
$\pi$. Nevertheless, there
does not exist a geodesic connecting $p$ with $r$.

\section{Spatial distance in the Szekeres model}

Now, let us turn to the physical point of view. To this end
we additionally require the validity of Einstein's equation;
for simplicity, we restrict ourselves to incoherent
matter (dust), i.e., to an ideal fluid with positive energy
density and vanishing pressure. Moreover, we require
the velocity--vector to be hypersurface--orthogonal.

One example of such space--times is due to Ellis [4];
its metric reads
\begin{equation}
ds^2 \ = \ dt^2 \ - \ t^{4/3}
 \{[b(x)/t + c(x)]^2 \,  dx^2 \, + \, dy^2 \, +  \, dz^2 \}
\end{equation}
with arbitrary positive functions $b$ and $c$.
It belongs to the Szekeres class [5]. The
velocity--vector is orthogonal
to the surfaces $[t \ = \ const.]$,
and the energy density $\rho$ equals
\begin{equation}
\rho \ = \ \frac{c}{6 \, \pi \, t \, (b \, + \, ct)}
\end{equation}
cf. [6] and [1]. Now the question arose how to define the
spatial distance within this model, if the events $p$ and
$q$ differ by their $x$--coordinate only.

In [1], it is argued that there is exactly one
connecting space--like geodesic, and its
length is most naturally considered as
distance between $p$ and $q$.

In [6], however, the following idea was implicitly made:
 According to Einstein, spatial distances have
to be measured by rigid bodies (rods), we consider the
rods to be at rest in comparison with the remaining
matter and put them together along a straightedge.
 In mathematical terms this means that we
consider the 3--dimensional Riemannian manifold
defined by $[t \ = \ const.]$ and take the
geodesic distance using the induced 3--metric.

So there is no contradiction between the calculations
made in [1] and [6] resp., but there have been
applied  different definitions of spatial
distances. It should be added that under a
 presumption  like $ distance \ << \ T \, \cdot \, c$
(where $T$ is the age of the universe and $c$
is the light velocity) both definitions
become approximately the same.

Near the singularity, however, both definitions
essentially differ, cf. [1]. If one prefers
the 4--geodesic version [1], one should
realize that for measuring the distance between
$p \ = \ (t_o, \, x_p, \,  y, \, z)$
 and
$q \ = \ (t_o, \, x_q, \,  y, \, z)$
one essentially needs the value of the metric
at earlier times $t < t_o$, whereas the 3--geodesic
version
[6] uses the metric at $t=t_0$ only.

So we think that at least for the model considered in
this section, the definition used in [6] closer
reflects what one wants to have as a definition
of spatial distances in General Relativity Theory.

\bigskip

{\Large
{\bf References}}

\noindent
[1] Griffiths, J. (1995). {\it Gen. Rel. Grav.}
{\bf 27}, 905.

\noindent
[2] Hawking, S., Ellis, G. F. R. (1973). {\it The large
scale structure of space--time}
 (Cambridge Univ. Press).

\noindent
[3] Schmidt, H.-J. (1993). {\it Fortschr. Phys.}
 {\bf 41}, 179.

\noindent
[4] Ellis, G. F. R. (1967). {\it J. Math. Phys.}
 {\bf 8}, 1171.

\noindent
[5] Szekeres, P. (1975). {\it Commun. Math. Phys.}
 {\bf 41}, 55.

\noindent
[6] Schmidt, H.-J. (1982). {\it Astron. Nachr.}
 {\bf 303}, 283.

\end{document}